
\magnification=1200
\baselineskip=13pt
\overfullrule=0pt
\tolerance=100000
{\hfill \hbox{\vbox{\settabs 1\columns
\+ UR-1367 \cr
\+ ER-40685-817\cr
\+ hep-th/9408049\cr}}}

\bigskip

\baselineskip=18pt

\centerline{\bf A Nonstandard Supersymmetric KP Hierarchy}

\vfill

\centerline{J.C. Brunelli}
\centerline{Department of Physics}
\centerline{University of Rochester}
\centerline{Rochester, NY 14627, USA}
\medskip
\centerline{and}
\medskip
\centerline{Ashok Das${}^*$}
\centerline{Instituto de F\'\i sica}
\centerline{Universidade Federal do Rio de Janeiro}
\centerline{Caixa Postal 68528}
\centerline{21945, Rio de Janeiro, Brasil}
\vfill

\centerline{\bf {Abstract}}

\medskip

We show that the supersymmetric nonlinear Schr\"odinger equation can be written
as a constrained super KP flow in a nonstandard representation of the Lax
equation. We construct the conserved charges and show that this system reduces
to the super mKdV equation with appropriate identifications. We construct
various flows generated by the general nonstandard super Lax equation and show
that they contain both the KP and mKP flows in the bosonic limits. This
nonstandard supersymmetric KP hierarchy allows
us to construct a new super KP equation which is nonlocal.
\vfill
\item{$*$}{Permanent address: Department of Physics and Astronomy, University
of Rochester, Rochester, NY 14627, USA.}

\eject

\noindent {\bf 1. {Introduction}}

\medskip

Integrable models have been studied vigorously in recent years from various
points of view [1-3]. In particular, we note that various two dimensional
gravity theories and continuum string equations arise naturally from the
study of such stystems. Even matrix models, in their continuum limit,
contain such systems and this has led to a lot of interest in their study
(see [4] and references therein).
These are models in $1+1$ or $2+1$ dimensions which are most
commonly represented in terms of a Lax operator which is a pseudo-differential
operator of the general form [5]
$$
L=\partial^n+u_{-1}\partial^{n-1}+u_0\partial^{n-2}+\cdots+u_{n-3}
\partial^{-1}+\cdots\eqno(1.1)
$$
Here $u_i(x)$'s are dynamical variables and their time evolution is given in
terms of a Lax equation which in the standard representation has the form
$$
{\partial L\over\partial t_{k}}=\left[L,(L^{k/n})_+\right]=
-\left[L,(L^{k/n})_-\right]\eqno(1.2)
$$
where  $(L^{k/n})_-$ ( $(L^{k/n})_+$ ) denotes the part of the
pseudo-differential operator containing only negative (nonnegative) powers of
$\partial$. Most integrable models studied in $1+1$ and $2+1$ dimensions have a
standard Lax representation of the form in  (1.2).

It is known, by now, that there exist integrable models which have a
nonstandard Lax representation of the form
$$
{\partial L\over\partial t_{k}}=
\left[L,(L^{k/n})_{\ge1}\right]\eqno(1.3)
$$
where $(\ \ )_{\ge1}$ represents the projection onto the purely differential
part of a pseudo-differential operator. The dispersive long water
wave equation [6]
or equivalently the two boson hierarchy [7-11]
has been studied from this point of
view and this in turn has led to the study of constrained KP hierarchies [12].
However, not much is known about the properties of the supersymmetric
generalizations of such system. In a recent paper [13], we studied the
supersymmetrization of the two boson hierarchy and showed how it gives the
supersymmetric nonlinear Schr\"odinger (NLS) equation [14,15]
with appropriate field
redefinitions. In the present paper we report on further general results in the
study of nonstandard supersymmetric Lax systems.

In sec. 2 we review briefly known results on the formulation of the nonlinear
Schr\"odinger equation as a constrained KP system with our observations that
become useful in the later sections. In sec. 3, we shown how the
supersymmetric nonlinear Schr\"odinger equation can be written as a constrained
super KP system but with a nonstandard Lax representation. We construct the
conserved charges and one of the Hamiltonian structures associated with this
system. We also show how the supersymmetric mKdV equation can be embedded into
this system with appropriate field identifications. In sec. 4 we study various
flows associated with a supersymmetric nonstandard KP system. We show that in
the bosonic limit, this system contains both the KP as well as the mKP flows.
This allows us to construct in sec. 5 a new supersymmetric KP equation which is
nonlocal. It, however, leads upon reduction to the supersymmetric KdV equation.
We present our conclusions in sec. 6.
\medskip

\noindent {\bf 2. {NSE As a Constrained KP System}}

\medskip

The two boson hierarchy is represented by a Lax operator of the form [6,11]
$$
L=\partial-J_0+\partial^{-1}J_1\eqno(2.1)
$$
and the nonstandard Lax equation
$$
{\partial L\over\partial t}=\left[L,\left(L^2\right)_{\ge1}\right]\eqno(2.2)
$$
leads to the system of integrable equations
$$
\eqalign{
{\partial J_0\over\partial t}=&(2J_1+J_0^2-J_0')'\cr
\noalign{\vskip 4pt}%
{\partial J_1\over\partial t}=&(2J_0J_1+J_1')'\cr
}\eqno(2.3)
$$
where a prime denotes differentiation with respect to $x$. It is now straight
forward to check that with the field identifications [7-10]
$$
\eqalign{
J_0=&-{q'\over q}=-(\ln q )'\cr
J_1=&{\bar q}q
}\eqno(2.4)
$$
the system of equations in $(2.3)$ reduce to the nonlinear Schr\"odinger
equation
$$
\eqalign{
{\partial q\over\partial t}=&-(q''+2({\bar q}q)q)\cr
\noalign{\vskip 4pt}%
{\partial {\bar q}\over\partial t}=&{\bar q}''+2({\bar q}q){\bar q}\cr
}\eqno(2.5)
$$

Let us next consider the Lax operator  (2.1) with the field
identifications in  (2.4) and note that
$$
\eqalign{
L=&\partial+{q'\over q}+\partial^{-1}{\bar q}q \cr
 =&q^{-1}(\partial+q\partial^{-1}{\bar q})q \cr
 =&G{\widetilde L}G^{-1}\cr
}\eqno(2.6)
$$
where
$$
\eqalign{
G=&q^{-1}\cr
{\widetilde L}=&\partial+q\partial^{-1}{\bar q}\cr
}\eqno(2.7)
$$
The two Lax operators, $L$ and ${\widetilde L}$, are said to be related through
a gauge transformation [7,10,16]. However, it can
be easily checked that in terms of the
Lax operator ${\widetilde L}$, the nonlinear Schr\"odinger equation can be
written in the standard Lax representation
$$
{\partial {\widetilde L}\over\partial t}=\left[{\widetilde L},
({\widetilde L}^2)_{+}\right]\eqno(2.8)
$$
Let us note that the Lax operator, $\widetilde L$, in (2.7) can also be
written as
$$
\eqalign{
{\widetilde L}=&\partial +q{\bar q}\,\partial^{-1}-q{\bar q}'\partial^{-2}+
q{\bar q}''\partial^{-3}+\cdots\cr
   =&\partial +\sum_{n=0}^\infty u_n\partial^{-n-1}\cr
}\eqno(2.9)
$$
with
$$
u_n=(-1)^nq{\bar q}^{(n)}\eqno(2.10)
$$
Here $f^{(n)}$ represents the $n$th derivative with respect to $x$. Note that
the form of $\widetilde L$ in the last expression in  (2.9) is the same as
that of a KP system. In this case, however, the coefficient functions are
constrained by  (2.10). Therefore, we can think of the nonlinear
Schr\"odinger equation as a constrained  KP system [7,9,12].

We will next make some observations on this system which will be useful in our
later discussions. First, let us note that given $\widetilde L$, we can define
its  formal adjoint [17]
$$
{\cal L}={\widetilde L}^*=-(\partial+{\bar q}\partial^{-1}q)\eqno(2.11)
$$
It is straight forward to check that the standard Lax equation
$$
{\partial{\cal L}\over\partial t}=\left[\left({\cal L}^2\right)_
+,{\cal L}\right]
\eqno(2.12)
$$
also gives the nonlinear Schr\"odinger equation. Furthermore, we note that with
the identification
$$
{\bar q}=q\eqno(2.13)
$$
the standard Lax equation
$$
{\partial {\widetilde L}\over\partial t}=\left[{\widetilde L},
({\widetilde L}^3)_{+}\right]\eqno(2.14)
$$
leads to the mKdV equation (the signs and factors can be appropriately
redefined by scaling of variables)
$$
{\partial q\over\partial t}=-(q'''+6q^2q')\eqno(2.15)
$$
The Lax operator, $\cal L$, with the identification in  (2.13) becomes
$$
{\cal L}=-{\widetilde L}\eqno(2.16)
$$
and also gives the mKdV equation as
$$
{\partial{\cal L}\over\partial t}=\left[\left({\cal L}^3\right)_+
,{\cal L}\right]
\eqno(2.17)
$$
This shows that the mKdV equation can be embedded into the nonlinear
Schr\"odinger equation and it appears from our discussion that the Lax operator
and its formal adjoint yield equivalent results.
\medskip

\noindent {\bf 3. {Super NSE As a Nonstandard Constrained Super KP
System}}

\medskip

We have shown in an earlier publication [13] that the supersymmetric two boson
hierarchy can be represented in the superspace by the Lax operator
$$
L = D^2 - (D \Phi_0) + D^{-1} \Phi_1\eqno(3.1)
$$
where $\Phi_0$ and $\Phi_1$ are two fermionic superfields and $D$ is the
covariant derivative in the superspace of the form
$$
D = {\partial  \over \partial  \theta} + \theta {\partial
 \over \partial  x}\eqno(3.2)
$$
The nonstandard Lax equation
$$
{\partial  L \over \partial  t} = \left[ L, \left( L^2 \right)_{\geq 1}
\right] \eqno(3.3)
$$
leads to the supersymmetric generalization of  (2.3), namely, (see ref. 12 for
details)
$$
\eqalign{
{\partial  \Phi_0 \over \partial  t} &=
 - (D^4 \Phi_0) + 2 (D \Phi_0)(D^2 \Phi_0) + 2(D^2 \Phi_1)\cr
\noalign{\vskip 4pt}%
{\partial  \Phi_1 \over \partial  t} &=
 (D^4 \Phi_1) + 2 \left(D^2((D \Phi_0) \Phi_1)\right)\cr
}\eqno(3.4)
$$
The system of equations (3.4) reduce to the supersymmetric nonlinear
Schr\"odinger equation of the form
$$
\eqalign{{
\partial  Q \over \partial  t} &=
 -(D^4 Q) + 2\left(D((DQ){\overline Q})\right)Q\cr
\noalign{\vskip 4pt}%
{\partial  \overline Q \over \partial  t} &=
 (D^4 \overline Q ) - 2\left(D((D{\overline Q})Q)\right){\overline Q}\cr
}\eqno(3.5)
$$
with the field identifications
$$
\eqalign{
\Phi_0 &= - D \ln (DQ) +
D^{-1} (\overline Q Q)\cr
\Phi_1 &= - \overline Q ( DQ)\cr}
\eqno(3.6)
$$
Here $Q$ and $\overline Q$ are fermionic superfields.

Let us next consider the Lax operator  (3.1) with the field
identifications in  (3.6) and note that
$$
\eqalign{
L=&D^2+{(D^3Q)\over(DQ)}-{\overline Q}Q-D^{-1}{\overline Q}(DQ)\cr
\noalign{\vskip 4pt}%
=&(DQ)^{-1}\left(D^2-{\overline Q}Q-(DQ)D^{-1}{\overline Q}\right)(DQ)\cr
=&G{\widetilde L}G^{-1}
}\eqno(3.7)
$$
where
$$
\eqalign{
G=&(DQ)^{-1}\cr
{\widetilde L}=&D^2-{\overline Q}Q-(DQ)D^{-1}{\overline Q} \cr
}\eqno(3.8)
$$
The two Lax operators, $L$ and $\widetilde L$, are related by a gauge
transformation in superspace. This is very much like the bosonic case. However,
unlike our earlier discussion, $\widetilde L$ does not lead to any consistent
equation in the standard or nonstandard representation of the Lax equation.

Let us next note that the formal adjoint of $\widetilde L$
in (3.8) can be written
as
$$
{\cal L}={\widetilde L}^*=-\left(D^2+{\overline Q}Q-{\overline Q}
D^{-1}(DQ)\right)\eqno(3.9)
$$
Through straight forward calculations, it can now be checked that the
nonstandard Lax equation
$$
{\partial{\cal L}\over\partial t}=\left[{\cal L},
\left({\cal L}^2\right)_{\ge1}\right]
\eqno(3.10)
$$
gives the supersymmetric nonlinear Schr\"odinger equations of (3.5). We see
that there are two basic differences from the bosonic case discussed in sec.
2. First, it is the formal
adjoint of the gauge transformed Lax operator which leads
to consistent equations and second, the supersymmetric generalization of
(2.5) is obtained as a nonstandard Lax equation.

We also note that we can write

$$
\eqalign{
{\cal L}=&-\left(D^2+{\overline Q}Q-{\overline Q}D^{-1}(DQ)\right)\cr
=&-\left(D^2+{\overline Q}Q- {\overline Q}(DQ)D^{-1}-{\overline Q}(D^2Q)D^{-2}
+{\overline Q}(D^3Q)D^{-3}+\cdots\right)\cr
=&-\left(D^2+\sum_{n=-1}^\infty\Phi_n D^{-n}\right)\cr
}\eqno(3.11)
$$
where
$$
\eqalign{
\Phi_{-1}=&0\cr
\Phi_n=&(-1)^{[{n+1\over2}]}\,{\overline Q}(D^nQ),\quad n\ge 0}
\eqno(3.12)
$$
Here $\Phi_{2n}$ ($\Phi_{2n+1}$) are bosonic (fermionic)  superfields and
$[n/2]$ stands for the integral part of $n/2$. Also, in (3.11) we have used the
generalized Liebnitz rule given in ref. 18. There
are several points to emphasize here. First, we note that the form of $\cal L$
in  (3.11) is identical to that of the Lax operator for susy KP. However, it
is an even parity Lax operator [19] and not of the Manin-Radul type [18]
(our result, therefore, differs from the one in ref. 20). Second,
the coefficient superfields $\Phi_n$ are
constrained by (3.12). And finally, the Lax equation that gives the
supersymmetric nonlinear Schr\"odinger equation, namely  (3.10), is a
nonstandard type. Therefore, we can think of the supersymmetric nonlinear
Schr\"odinger equation as a nonstandard, constrained supersymmetric KP system.

{}From the structure of the Lax equation in  (3.10), one can show that the
conserved quantities of the system are given by
$$
H^{(n)}={1\over n}\int dx d\theta\,\hbox{sRes}\,{\cal L}^n={1\over n}\int
d\mu\,
\hbox{sRes}\,{\cal L}^n\eqno(3.13)
$$
where ``sRes'' stands for the super residue which is defined to be the
coefficient of $D^{-1}$ ($D^{-1}$ is assumed to be on the right). The first few
conserved quantities have the form
$$
\eqalign{
H^{(1)}=&\int d\mu\,{\overline Q}(DQ)\cr
\noalign{\vskip 4pt}%
H^{(2)}=&{1\over 2}\int d\mu\,\left({\overline Q}(D^3Q)-(D^3{\overline Q})Q
\right)\cr
\noalign{\vskip 4pt}%
H^{(3)}=&-{1\over 2}\int d\mu\,\Bigl[(D^3{\overline Q})(D^2Q)+
(D^2{\overline Q})(D^3Q)\Bigr.\cr
&\phantom{-{1\over 2}\int d\mu\,}+(D{\overline Q})(DQ)
\left((D{\overline Q})Q+{\overline Q}(DQ)\right)\cr
&\phantom{-{1\over 2}\int d\mu\,}-{\overline Q}Q
\Bigr.\left((D^2{\overline Q})(DQ)-(D{\overline Q})(D^2Q)\right)\Bigl]\cr
}\eqno(3.14)
$$
These can be compared with the conserved quantities of ref. 14 for $k=1$. It
can
also be checked that the system of equations (3.5) are Hamiltonian with respect
to $H^{(2)}$ and the Hamiltonian structure
$$
\eqalign{
\{Q(x_1,\theta_1,t),Q(x_2,\theta_2,t)\}=&Q(x_1,\theta_1,t)Q(x_2,\theta_2,t)
D_1^{-1}\Delta_{12}\cr
\noalign{\vskip 4pt}%
\{Q(x_1,\theta_1,t),{\overline Q}(x_2,\theta_2,t)\}=&-{1\over2}D_1\Delta_{12}-
Q(x_1,\theta_1,t){\overline Q}(x_2,\theta_2,t)
D_1^{-1}\Delta_{12}\cr
\noalign{\vskip 4pt}%
\{{\overline Q}(x_1,\theta_1,t),{\overline Q}(x_2,\theta_2,t)\}=
&{\overline Q}(x_1,\theta_1,t){\overline Q}(x_2,\theta_2,t)
D_1^{-1}\Delta_{12}\cr
}\eqno(3.15)
$$
where
$$
\Delta_{12}=\delta(x_1-x_2)\delta(\theta_1-\theta_2)\eqno(3.16)
$$

To conclude this section, let us note that if we identify
$$
{\overline Q}=Q\eqno(3.17)
$$
then
$$
{\cal L}=-\left(D^2-QD^{-1}(DQ)\right)\eqno(3.18)
$$
and unlike the bosonic case, it is different from its formal
adjoint with the same
identification. It can also be checked that with the identification in
(3.17), the nonstandard Lax equation
$$
{\partial{\cal L}\over\partial t}=\left[{\cal L},
\left({\cal L}^3\right)_{\ge1}\right]
\eqno(3.19)
$$
yields
$$
{\partial  Q \over \partial  t} =-(D^6 Q) +3\left(D^2(Q(DQ))\right)(DQ)
\eqno(3.20)
$$
This is nothing other than the supersymmetric mKdV equation [21]
and this shows how
the susy mKdV equation can be embedded into the susy nonlinear Schr\"odinger
equation in a nonstandard Lax representation. This is quite analogous to the
embedding of the supersymmetric KdV equation in the supersymmetric two boson
hierarchy (see ref. 13 for details).
\medskip
\noindent {\bf 4. {General Flows of the Nonstandard Super KP
Hierarchy}}

\medskip

Let us consider a general super Lax operator of the form (3.11)
$$
\eqalign{
L&=D^2+\Phi_0+\Phi_1D^{-1}+\Phi_2D^{-2}+\cdots\cr
 &=D^2+\sum_{n=0}^\infty\Phi_nD^{-n}
}\eqno(4.1)
$$
where the Grassmann parity of the superfields $\Phi_n$ are
$$
|\Phi_n|={1-(-1)^n\over 2}\eqno(4.2)
$$
Let the expansion of the superfields be of the form
$$
\eqalign{
 \Phi_{2n}=&q_{2n}+\theta \phi_{2n}\cr
 \Phi_{2n+1}=&\phi_{2n+1}+\theta q_{2n+1}\cr
}\eqno(4.3)
$$
where $q_n$ ($\phi_n$) are the bosonic (fermionic) components of the
superfields.

The nonstandard flows associated with this super KP Lax operator are given by
$$
{\partial L\over\partial t_n}=\left[\left(L^n\right)_{\ge1},L\right]\eqno(4.4)
$$
For $n=1$, the flow is quite trivial and gives
$$
{\partial\Phi_n\over\partial t_1}=
(D^2\Phi_n)=\left({\partial\Phi_n\over\partial x}\right)\eqno(4.5)
$$
This implies that the  time coordinate $t_1$ can be identified with $x$.

For $n=2$, the flow in  (4.4) gives
$$
\eqalign{
{\partial\Phi_n\over\partial t_2}=&
(D^4\Phi_n)+2(D^2\Phi_{n+2})+2\Phi_0(D^2\Phi_n)
+2\Phi_1(D\Phi_n)-2(1+(-1)^{n})\Phi_1\Phi_{n+1}\cr
&+2\sum_{\ell\ge1}\left\{-(-1)^{[{\ell/2}]}\left[\matrix{n+1\cr \ell}
\right]\Phi_{n-\ell+2}(D^{\ell}\Phi_0)
+(-1)^{[{\ell/2}]+n}\left[\matrix{n\cr \ell}\right]
\Phi_{n-\ell+1}(D^{\ell}\Phi_1)\right\}
}\eqno(4.6)
$$
where the super binomial coefficients $\left[\matrix{n\cr \ell}\right]$  are
defined in ref. 18. The equations for the bosonic components can be obtained
from  (4.6) to be
$$
\eqalign{
{\partial q_{2n}\over\partial
t_2}=&q_{2n}''+2q_{2n+2}'+2q_0q_{2n}'+2\phi_1\phi_{2n}-4\phi_1\phi_{2n+1}\cr
&+2\sum_{\ell\ge1}(-1)^{\ell}\Biggl\{-\left[\matrix{2n+1\cr 2\ell}\right]
q_{2n-2\ell+2}\,q_0^{(\ell)}\Biggr.+\left[\matrix{2n+1\cr 2\ell-1}\right]
\phi_{2n-2\ell+3}\,\phi_0^{(\ell-1)}\cr
&\phantom{+2\sum_{\ell\ge1}(-1)^{\ell}\Biggl\{ }
+\left[\matrix{2n\cr 2\ell}\right]
\phi_{2n-2\ell+1}\,\phi_1^{(\ell)}-\left[\matrix{2n\cr 2\ell-1}\right]
q_{2n-2\ell+2}\,q_1^{(\ell-1)}\Biggl.\Biggr\}
}\eqno(4.7)
$$
$$
\eqalign{
{\partial q_{2n+1}\over\partial
t_2}=&q_{2n+1}''+2q_{2n+3}'+2(\phi_0\phi_{2n+1}'+q_0q_{2n+1}')
+2(q_1q_{2n+1}-\phi_1\phi_{2n+1}')\cr
&+2\sum_{\ell\ge1}(-1)^{\ell}\Biggl\{-\left[\matrix{2n+2\cr 2\ell}\right]
(-\phi_{2n-2\ell+3}\,\phi^{(\ell)}_0+q_{2n-2\ell+3}\,q_0^{(\ell)})\Biggr.\cr
&\phantom{+2\sum_{\ell\ge1}(-1)^{\ell}\Biggl\{ }
+\left[\matrix{2n+2\cr 2\ell-1}\right]
(\phi_{2n-2\ell+4}\,\phi_0^{(\ell-1)}+q_{2n-2\ell+4}\,q_0^{(\ell)})\cr
&\phantom{+2\sum_{\ell\ge1}(-1)^{\ell}\Biggl\{ }
-\left[\matrix{2n+1\cr 2\ell}\right]
(\phi_{2n-2\ell+2}\,\phi_1^{(\ell)}+q_{2n-2\ell+2}\,q_1^{(\ell)})\cr
&\phantom{+2\sum_{\ell\ge1}(-1)^{\ell}\Biggl\{ }
+\left[\matrix{2n+1\cr 2\ell-1}\right]
(q_{2n-2\ell+3}\,q_1^{(\ell-1)}-\phi_{2n-2\ell+3}\,\phi_1^{(\ell)})
\Biggl.\Biggr\}
}\eqno(4.8)
$$
\eject
\noindent In the bosonic limit -- when all
the $\phi_n$'s are zero -- we note that if we set

$$
q_{2n}=0\,,\quad\hbox{for all }n\eqno(4.9)
$$
and identify
$$
q_{2n+1}=u_n\,,\quad\hbox{for all }n\eqno(4.10)
$$
then  (4.8) gives
$$
\eqalign{
{\partial u_0\over\partial t_2}=&u_0''+2u_1'\cr
{\partial u_1\over\partial t_2}=&u_1''+2u_2'+2u_0u_0'\cr
{\partial u_2\over\partial t_2}=&u_2''+2u_3'-2u_0u_0''+4u_1u_0'\cr
\vdots&\cr
}\eqno(4.11)
$$
which are nothing other than the $t_2$-flows for the standard KP hierarchy
[16,22].

On the other hand, in the bosonic limit, if we set
$$
q_{2n+1}=0\,,\quad\hbox{for all }n\eqno(4.12)
$$
and identify
$$
q_{2n}=u_n\,,\quad\hbox{for all }n\eqno(4.13)
$$
then, (4.7) gives
$$
\eqalign{
{\partial u_0\over\partial t_2}=&u_0''+2u_1'+2u_0u_0'\cr
{\partial u_1\over\partial t_2}=&u_1''+2u_2'+2u_0u_1'+2u_0'u_1\cr
{\partial u_2\over\partial t_2}=&u_2''+2u_3'-2u_1u_0''+2u_0u_2'+4u_2u_0'\cr
\vdots&\cr
}\eqno(4.14)
$$
which are nothing other than the $t_2$-flows associated with the mKP
hierarchy [16,22].

For $n=3$, equation (4.4) gives
\eject
$$
\eqalign{
{\partial\Phi_n\over\partial t_3}=&
(D^6\Phi_n)+3(D^4\Phi_{n+2})+3(D^2\Phi_{n+4})
+3\Phi_0(D^4\Phi_n)+6\Phi_0(D^2\Phi_{n+2})\cr
&+3\Phi_1(D^3\Phi_n)+3\Phi_1(D\Phi_{n+2})-3(-1)^n\Phi_1(D^2\Phi_{n+1})\cr
&-3(1+(-1)^n)\Phi_1\Phi_{n+3}-
3(1+(-1)^n)((D^2\Phi_1)+\Phi_3+2\Phi_1\Phi_0)\Phi_{n+1}\cr
&+3((D^2\Phi_0)+\Phi_2+\Phi_0^2)(D^2\Phi_n)
+3((D^2\Phi_1)+\Phi_3+2\Phi_1\Phi_0)(D\Phi_n)\cr
&+3\sum_{\ell\ge1}\Biggl\{-(-1)^{[\ell/2]}
\left[\matrix{n+3\cr \ell}\right]\Phi_{n-\ell+4}(D^{\ell}\Phi_0)\cr
\noalign{\vskip -2pt}%
&\phantom{+3\sum_{\ell\ge1}\Biggl\{}+(-1)^{[{\ell/2}]+n}
\left[\matrix{n+2\cr \ell}\right]\Phi_{n-\ell+3}(D^{\ell}\Phi_1)\cr
\noalign{\vskip -2pt}%
&\phantom{+3\sum_{\ell\ge1}\Biggl\{}+(-1)^{[{\ell/2}]}
\left[\matrix{n+1\cr \ell}\right]\Phi_{n-\ell+2}(D^{\ell}
\left((D^2\Phi_0)+\Phi_2+\Phi_0^2)\right)\cr
\noalign{\vskip -2pt}%
&\phantom{+3\sum_{\ell\ge1}\Biggl\{}+(-1)^{[{\ell/2}]+n}
\left[\matrix{n\cr \ell}\right]\Phi_{n-\ell+1}(D^{\ell}
\left((D^2\Phi_1)+\Phi_3+2\Phi_1\Phi_0)\right)\Biggl.\Biggr\}
}\eqno(4.15)
$$
The bosonic components can again be obtained from  (4.15)
and they have the form
$$
\eqalign{
{\partial q_{2n}\over\partial t_3}=&q_{2n}'''+3q_{2n+2}''+3q_{2n+4}'+
3q_0q_{2n}''+6q_0q_{2n+2}'
+3\phi_1\phi_{2n}'\cr
&+3\phi_1\phi_{2n+2}-3\phi_1\phi_{2n+1}'-6\phi_1\phi_{2n+3}
-6(\phi_1'+2q_0\phi_1+\phi_3)\phi_{2n+1}\cr
&+3(q_0^2+q_0'+q_2)q_{2n}'+3(\phi_1'+2q_0\phi_1+\phi_3)\phi_{2n}\cr
&+3\sum_{\ell\ge1}(-1)^{\ell}\Biggl\{-\left[\matrix{2n+3\cr 2\ell}\right]
q_{2n-2\ell+4}\,q_0^{(\ell)}\Biggr.+\left[\matrix{2n+3\cr 2\ell-1}\right]
\phi_{2n-2\ell+5}\,\phi_0^{(\ell-1)}\cr
\noalign{\vskip -2pt}%
&\phantom{+3\sum_{\ell\ge1}(-1)^{\ell}\Biggl\{ }
+\left[\matrix{2n+2\cr 2\ell}\right]
\phi_{2n-2\ell+3}\,\phi_1^{(\ell)}-\left[\matrix{2n+2\cr 2\ell-1}\right]
q_{2n-2\ell+4}\,q_1^{(\ell-1)}\cr
\noalign{\vskip -2pt}%
&\phantom{+3\sum_{\ell\ge1}(-1)^{\ell}\Biggl\{ }
-\left[\matrix{2n+1\cr 2\ell}\right]
q_{2n-2\ell+2}(q_0^2+q_0'+q_2)^{(\ell)}\cr
\noalign{\vskip -2pt}%
&\phantom{+3\sum_{\ell\ge1}(-1)^{\ell}\Biggl\{ }
+\left[\matrix{2n+1\cr 2\ell-1}\right]
\phi_{2n-2\ell+3}(2q_0\phi_0+\phi_0'+\phi_2)^{(\ell-1)}\cr
\noalign{\vskip -2pt}%
&\phantom{+3\sum_{\ell\ge1}(-1)^{\ell}\Biggl\{ }
+\left[\matrix{2n\cr 2\ell}\right]
\phi_{2n-2\ell+1}(\phi_1'+2q_0\phi_1+\phi_3)^{(\ell)}\cr
\noalign{\vskip -2pt}%
&\phantom{+3\sum_{\ell\ge1}(-1)^{\ell}\Biggl\{ }
-\left[\matrix{2n\cr 2\ell-1}\right]
q_{2n-2\ell+2}(q_1'+2q_0q_1+2\phi_0\phi_1+q_3)^{(\ell-1)}
\Biggl.\Biggr\}
}\eqno(4.16)
$$
\eject
$$
\eqalign{
{\partial q_{2n+1}\over\partial
t_3}=&q_{2n+1}'''+3q_{2n+3}''+3q_{2n+5}'+3(q_0q_{2n+1}''+\phi_0\phi_{2n+1}'')\cr
&+6(q_0q_{2n+3}'+\phi_0\phi_{2n+3}')+3(q_1q_{2n+1}''-\phi_1\phi_{2n+1}'')\cr
&+3(q_1q_{2n+3}-\phi_1\phi_{2n+3}')+3(q_1q_{2n+2}'-\phi_1\phi_{2n+2}')\cr
&+3(2q_0\phi_0+\phi_0'+\phi_2)\phi_{2n+1}'+3(q_0^2+q_0'+q_2)q_{2n+1}'\cr
&+3(q_1'+2q_0q_1+2\phi_0\phi_1+q_3)q_{2n+1}-3(\phi_1'+2q_0\phi_1+2q_1\phi_0+
\phi_3)\phi_{2n+1}'\cr
&+3\sum_{\ell\ge1}(-1)^{\ell}\Biggl\{-\left[\matrix{2n+4\cr 2\ell}\right]
(q_{2n-2\ell+5}\,q_0^{(\ell)}-\phi_{2n-2\ell+5}\,\phi_0^{(\ell)})\Biggr.\cr
&\phantom{+3\sum_{\ell\ge1}(-1)^{\ell}\Biggl\{ }
+\left[\matrix{2n+4\cr 2\ell-1}\right]
(\phi_{2n-2\ell+6}\,\phi_0^{(\ell-1)}+q_{2n-2\ell+6}\,q_0^{(\ell)})\cr
&\phantom{+3\sum_{\ell\ge1}(-1)^{\ell}\Biggl\{ }
-\left[\matrix{2n+3\cr 2\ell}\right]
(\phi_{2n-2\ell+4}\,\phi_1^{(\ell)}+q_{2n-2\ell+4}\,q_1^{(\ell)})\cr
&\phantom{+3\sum_{\ell\ge1}(-1)^{\ell}\Biggl\{ }
+\left[\matrix{2n+3\cr 2\ell-1}\right]
(q_{2n-2\ell+5}\,q_1^{(\ell-1)}-\phi_{2n-2\ell+5}\,\phi_1^{(\ell)})\cr
&\phantom{+3\sum_{\ell\ge1}(-1)^{\ell}\Biggl\{ }
-\left[\matrix{2n+2\cr 2\ell}\right]
\left(q_{2n-2\ell+3}(q_0^2+q_0'+q_2)^{(\ell)}\right.\cr
\noalign{\vskip -8pt}%
&\hskip7.0truecm\left.-\phi_{2n-2\ell+3}(2q_0\phi_0+\phi_0'+\phi_2)^{(\ell)}
\right)\cr
\noalign{\vskip 5pt}%
&\phantom{+3\sum_{\ell\ge1}(-1)^{\ell}\Biggl\{ }
+\left[\matrix{2n+2\cr 2\ell-1}\right]
\left(\phi_{2n-2\ell+4}(2q_0\phi_0+\phi_0'+\phi_2)^{(\ell-1)}\right.\cr
\noalign{\vskip -8pt}%
&\hskip7.0truecm\left.+q_{2n-2\ell+4}(q_0^2+q_0'+q_2)^{(\ell)}\right)\cr
\noalign{\vskip 5pt}%
&\phantom{+3\sum_{\ell\ge1}(-1)^{\ell}\Biggl\{ }
-\left[\matrix{2n+1\cr 2\ell}\right]
\left(\phi_{2n-2\ell+2}(\phi_1'+2q_0\phi_1+2q_1\phi_0+\phi_3)^{(\ell)}\right.\cr
\noalign{\vskip -8pt}%
&\hskip6.0truecm\left.+q_{2n-2\ell+2}(q_1'+2q_0q_1+2\phi_0\phi_1+q_3)^{(\ell)}
\right)\cr
\noalign{\vskip 5pt}%
&\phantom{+3\sum_{\ell\ge1}(-1)^{\ell}\Biggl\{ }
+\left[\matrix{2n+1\cr 2\ell-1}\right]
\left(q_{2n-2\ell+3}(q_1'+2q_0q_1+2\phi_0\phi_1+q_3)^{(\ell-1)}\right.\cr
\noalign{\vskip -14pt}%
&\hskip6.0truecm
\left.-\phi_{2n-2\ell+3}(\phi_1'+2q_0\phi_1+2q_1\phi_0+\phi_3)^{(\ell)}\right)
\Biggl.\Biggr\}
}\eqno(4.17)
$$

\noindent  We note here for
completeness that, in the bosonic limit, with the
identifications in  (4.9) and (4.10), we obtain from (4.17)
\vfill\eject
$$
\eqalign{
{\partial u_0\over\partial t_3}=&u_0'''+3u_1''+3u_2'+6u_0u_0'\cr
{\partial u_1\over\partial t_3}=&u_1'''+3u_2''+3u_3'+6u_0u_1'+6u_0'u_1\cr
\vdots&\cr
}\eqno(4.18)
$$
which are the $t_3$-flows for the standard KP hierarchy [16,22].
On the other hand, the
identifications in  (4.12) and (4.13) lead to (from  (4.16))
$$
\eqalign{
{\partial u_0\over\partial t_3}=&u_0'''+3u_1''+3u_2'+3u_0u_0''+3(u_0')^2+
6(u_1u_0)'+3u_0^2u_0'\cr
\vdots&\cr
}\eqno(4.19)
$$
These are  the $t_3$-flows for the mKP hierarchy [16,22]. It is
interesting to note that the nonstandard KP equation (4.4) contains both the
standard KP and the mKP flows in its bosonic limit.
\medskip

\noindent {\bf 5. {A New Super KP Equation}}

\medskip

As we have shown in the last section, the  nonstandard KP equation of (4.4)
reduces to the standard KP flows in the bosonic limit. It is, therefore,
interesting to examine in some detail the nature of the super KP equation that
it leads to. To that end, we assume
$$
\Phi_{2n}=0\,,\quad\hbox{for all }n\eqno(5.1)
$$
The first two nontrivial equations following from  (4.6) with this
identifications are
$$
\eqalign{
{\partial\Phi_1\over\partial t_2}=&(D^4\Phi_1)+2(D^2\Phi_3)\cr
\noalign{\vskip 4pt}%
{\partial\Phi_3\over\partial t_2}=&(D^4\Phi_3)+2(D^2\Phi_5)-2(D(\Phi_1\Phi_3))
+2\Phi_1(D^3\Phi_1)\cr
}\eqno(5.2)
$$
Similarly, the first nontrivial equation following from (4.15) with the
identification in  (5.1) has the form
$$
{\partial\Phi_1\over\partial t_3}=(D^6\Phi_1)+3(D^4\Phi_3)+3(D^2\Phi_5)
+3(D^2(\Phi_1(D\Phi_1)))\eqno(5.3)
$$
{}From  (5.2) and (5.3), we obtain
$$
D^2\left( {\partial\Phi_1\over\partial t_3}-{1\over 4}(D^6\Phi_1)-
{3\over2}(D^2(\Phi_1(D\Phi_1)))-
{3\over2}( D ( \Phi_1 ( D^{-2}{\partial\Phi_1\over\partial t_2}
)))\right)={3\over 4}{\partial^2\Phi_1\over\partial t_2^2}
\eqno(5.4)
$$
With the identifications
$$
t_2=y,\ t_3=t \quad \hbox{and}\quad \Phi_1=\Phi=\phi+\theta u \eqno(5.5)
$$
equation (5.4) becomes
$$
D^2\left({\partial\Phi\over\partial t}-{1\over 4}(D^6\Phi)-
{3\over2}(D^2(\Phi(D\Phi)))-
{3\over2}(D(\Phi(\partial^{-1}{\partial\Phi\over\partial y})))\right)
={3\over 4}{\partial^2\Phi\over\partial y^2}\eqno(5.6)
$$

It is now straight forward to check that (5.6) reduces in the bosonic limit
to
$$
{\partial\ \over\partial x}\left({\partial u\over\partial t}-{1\over4}u'''
-3uu'\right)
={3\over4}{\partial^2 u\over\partial y^2}
\eqno(5.7)
$$
which is the KP equation. The supersymmetric generalization in  (5.6),
however, differs from the Manin-Radul equation [18,23]
because of the presence of the
nonlocal terms. We note that in components  (5.6) takes the form
$$
\eqalign{
{\partial\ \over\partial x}\left({\partial u\over\partial t}-{1\over4}u'''
-3uu'+{3\over2}\phi\phi''
-{3\over2}\phi'(\partial^{-1}{\partial\phi\over\partial y})
-{3\over2}\phi{\partial\phi\over\partial y}\right)
=&{3\over4}{\partial^2 u\over\partial y^2}\cr
\noalign{\vskip 5pt}%
{\partial\ \over\partial x}\left({\partial\phi\over\partial t}
-{1\over4}\phi'''-{3\over2}(u\phi)'
-{3\over2}u(\partial^{-1}{\partial\phi\over\partial y})+
{3\over2}\phi(\partial^{-1}{\partial u\over\partial y})
\right)
=&{3\over4}{\partial^2 \phi\over\partial y^2}\cr
}\eqno(5.8)
$$
These equations are not invariant under
$$
y\leftrightarrow-y\eqno(5.9)
$$
unlike the Manin-Radul equations. However, we note that when we restric
the variables $u$ and $\phi$  to be independent of $y$, these
equations reduce to the supersymmetric KdV equation [21].
These, therefore, represent  a new supersymmetric
generalization of the KP equation.

\medskip

\noindent {\bf 6. {Conclusion}}

\medskip

We have  shown that the supersymmetric nonlinear Schr\"odinger equation  can be
represented as a nonstandard, constrained super KP flow. We have constructed
the conserved quantities of the system in this formalism and we have shown how
the supersymmetric mKdV equation can be embedded into this system. We have
worked out the first three flows associated with a general, nonstandard, super
KP system and we have shown that these flows contain both the standard KP flows
as well as the mKP flows in their bosonic limit. We have shown that these flows
lead to a new supersymmetrization of the KP equation that is nonlocal. It has
the correct bosonic limit and when properly restricted, it reduces to the
supersymmetric KdV equation. However, this equation is different from the
Manin-Radul equation because of nonlocal terms which are also antisymmetric
under $y\leftrightarrow -y$. Properties of this system are under study and will
be reported in a later publication.

\medskip

\noindent {\bf Acknowledgements}

\medskip

This work was supported in part by the U.S. Department of Energy Grant No.
DE-FG-02-91ER40685.  We would also like to thank CNPq, Brazil, for
financial support.

\vfill\eject

\noindent {\bf {References}}

\medskip

\item{1.} L.D. Faddeev and L.A. Takhtajan, ``Hamiltonian Methods in
the Theory of Solitons'' (Springer, Berlin, 1987).

\item{2.} A. Das, ``Integrable Models'' (World Scientific, Singapore,
1989).

\item{3.} M.J. Ablowitz and P.A. Clarkson, ``Solitons, Nonlinear
Evolution Equations and Inverse Scattering'' (Cambridge, New York, 1991).

\item{4.} D. J. Gross and A. A. Midgal, Phys. Rev. Lett. {\bf 64} (1990) 127;
D. J. Gross and A. A. Midgal, Nucl. Phys. {\bf B340} (1990) 333; E. Br\'ezin
and V. A. Kazakov, Phys. Lett. {236B} (1990) 144; M. Douglas and S. H. Shenker,
Nucl. Phys. {\bf B335} (1990) 635;
A. M. Polyakov in ``Fields, Strings and Critical Phenomena'', Les
Houches 1988, ed. E. Br\'ezin and J. Zinn-Justin (North-Holland, Amsterdam,
1989); L. Alvarez-Gaum\'e, Helv. Phys. Acta {\bf 64} (1991) 361; P. Ginsparg
and G. Moore, ``Lectures on 2D String Theory and 2D Gravity'' (Cambridge, New
York, 1993).

\item{5.} L. A. Dickey, ``Soliton Equations and Hamiltonian Systems'' (World
Scientific, Singapore, 1991).

\item{6.} B.A. Kupershmidt, Commun. Math. Phys. {\bf 99} (1985) 51.

\item{7.} H. Aratyn, L.A. Ferreira, J.F. Gomes and A.H. Zimerman, Nucl.
 Phys. {\bf B402} (1993) 85; H. Aratyn, L.A. Ferreira, J.F. Gomes and A.H.
Zimerman, ``Lectures at the VII J. A. Swieca Summer School'', January 1993,
hep-th/9304152; H. Aratyn, E. Nissimov and S. Pacheva, Phys. Lett. {\bf B314}
(1993) 41.

\item{8.} L. Bonora and C.S. Xiong, Phys. Lett. {\bf B285} (1992) 191; L.
Bonora and C.S. Xiong, Int. J. Mod. Phys. {\bf A8} (1993) 2973.

\item{9.} M. Freeman and P. West, Phys. Lett. {\bf 295B} (1992) 59.

\item{10.} J. Schiff, ``The Nonlinear Schr\"odinger Equation and
Conserved Quantities in the Deformed Parafermion and SL(2,{\bf R})/U(1)
Coset Models'', Princeton preprint IASSNS-HEP-92/57 (1992)
(also hep-th/9210029).

\item{11.} J. C. Brunelli, A. Das and W.-J. Huang, Mod. Phys. Lett. {\bf 9A}
(1994) 2147.

\item{12.} W. Oevel and W. Strampp, Commun. Math. Phys. {\bf 157} (1993) 51,
and references therein.

\item{13.} J.C. Brunelli and A. Das, ``The Supersymmetric Two
Boson Hierarchies'',
University of Rochester preprint UR-1362 (1994) (also hep-th/9406214).

\item{14.} G.H.M. Roelofs and P.H.M. Kersten, J. Math. Phys. {\bf 33}
(1992) 2185.

\item{15.} J.C. Brunelli and A. Das, ``Tests of Integrability of the
Supersymmetric Nonlinear Schr\"odinger Equation'', University of Rochester
preprint UR-1344 (1994) (also hep-th/9403019).

\item{16.} W. Oevel and C. Rogers, Rev. Math. Phys. {\bf 5} (1993) 299.

\item{17.} E. Date, M. Kashiwara, M. Jimbo and T. Miwa, in ``Nonlinear
Integrable Systems-Classical Theory and Quantum Theory'', ed. M. Jimbo and T.
Miwa (World Scientific, Singapore, 1983).

\item{18.} Y. I. Manin and A. O. Radul, Commun. Math. Phys. {\bf 98} (1985) 65.

\item{19.} J. M. Figueroa-O'Farrill, J. Mas and E. Ramos, Rev. Math. Phys.
{\bf 3} (1991) 479; F. Yu, Nucl. Phys. {B375} (1992) 173; J. Barcelos-Neto, S.
Ghosh and S. Roy, ``The Hamiltonian Structures of the Super KP Hierarchy
Associated with an Even Parity SuperLax Operator'', ICTP preprint IC/93/179
(1993) (also hep-th/9307119).

\item{20.} F. Toppan, ``$N$=1,2 Super-NLS Hierarchies as Super-KP Coset
Reductions'', preprint ENSLAPP-L-467/94 (1994) (also hep-th/940595).

\item{21.} P. Mathieu, J. Math. Phys. {\bf 29} (1988) 2499.

\item{22.} Y. Ohta, J. Satsuma, D. Takahashi and T. Tokihiro, Progr. Theor.
Phys. Suppl. {\bf 94} (1988) 210; K. Kiso, Progr. Theor. Phys. {\bf 83}
(1990) 1108.

\item{23.} J. Barcelos-Neto, A. Das, S. Panda and S. Roy, Phys. Lett. {\bf
B282}
(1992) 365.

\end